\documentclass[aps,prapplied,preprint,superscriptaddress,amsmath,amssymb]{revtex4-2}
\usepackage[T1]{fontenc}
\usepackage[utf8]{inputenc}
\usepackage{graphicx}
\usepackage{hyperref}
\usepackage{xcolor}
\usepackage{soul}
\newcommand{\SI}[2]{#1\,\mathrm{#2}}

\begin{document}

\title{Direct Generation of Radially Polarized Vector Vortex Beam from an Exciton-Polariton Laser}
\author{Jiaqi Hu}
 \affiliation{Applied Physics Program, University of Michigan, Ann Arbor, MI 48109, USA}
\author{Seonghoon Kim}
 \affiliation{Department of Electrical Engineering, University of Michigan, Ann Arbor, MI 48109, USA}
\author{Christian Schneider}
 \affiliation{Institute of Physics, University of Oldenburg, D-26129 Oldenburg, Germany}
 \affiliation{Technische Physik, Universität Würzburg, Am Hubland, Würzburg 97074, Germany}
\author{Sven Höfling}
 \affiliation{Technische Physik, Universität Würzburg, Am Hubland, Würzburg 97074, Germany}
 \affiliation{SUPA, School of Physics and Astronomy, University of St Andrews, St Andrews KY16 9SS, United Kingdom}
\author{Hui Deng}
 \email{dengh@umich.edu}
 \affiliation{Applied Physics Program, University of Michigan, Ann Arbor, MI 48109, USA}
 \affiliation{Department of Physics, University of Michigan, Ann Arbor, MI 48109, USA}
\date{\today}

\begin{abstract}
Vector vortex beams are a class of optical beams with singularities in their space-variant polarization. 
Vector vortex beam lasers have applications in many areas including imaging and communication, where vertical cavity lasers emitting Gaussian beams have been most widely used so far. 
Generation of vector vortex beams from vertical cavity lasers has required external control or modulation. 
Here, by utilizing a polarization-selective sub-wavelength grating as one of the reflectors in a vertical semiconductor microcavity, we design the spin textures of the polariton mode and demonstrate polariton lasing in a single-mode, radially polarized vector vortex beam. 
Polarization and phase distributions of the emission are characterized by polarization-resolved imaging and interferometry. 
This way of vector vortex laser beam generation allows low threshold power, stable single-mode operation, scalability and on-chip integration, all of which are important for applications in imaging and communication. 
\end{abstract}

\maketitle

\section{Introduction}

Optical beams with non-uniform spatial distribution of polarization open up many possibilities due to this additional degree of freedom compared to conventional scalar beams with uniform polarization \cite{gori_polarization_2001, zhan_cylindrical_2009, rosales-guzman_review_2018}. 
In particular, vector vortex beams, identified by rotation of its polarization around a singularity point, have a broad range of potential applications ranging from high-resolution imaging \cite{quabis_focusing_2000, dorn_sharper_2003, chen_imaging_2013, segawa_demonstration_2014}, to optical manipulation \cite{kozawa_optical_2010, roxworthy_optical_2010} and optical communication \cite{ndagano_fiber_2015, ndagano_creation_2018}. 

Vector vortex beams can be generated by passing a free-space Gaussian laser beam or guided waves through additional polarization or phase modulation components \cite{bomzon_radially_2002, kozawa_generation_2005, phelan_generation_2011, cai_integrated_2012, arbabi_dielectric_2015}, integrating polarization and phase control components with a fiber laser \cite{zheng_all-fiber_2010} or distributed-feedback semiconductor laser \cite{zhang_inp-based_2018}, or introducing external frequency-selective feedback and anisotropy to vertical-cavity surface-emitting lasers (VCSELs) \cite{jimenez-garcia_spontaneous_2017}. 
These systems are often bulky, complex and difficult to control. 
An on-chip vector vortex beam laser, without the need of external control components, has the advantages of higher efficiency, more robustness, and scalable integration. 
It has been demonstrated recently in micro-ring lasers \cite{miao_orbital_2016}. 
In the more widely used vertical cavity architecture, a weak-coupling regime VCSEL with metallic gratings on top of the output surface was shown to emit a vector vortex beam, but the polarization-selective transmission depended on lossy surface plasmon polariton modes \cite{cai_generation_2010}. 
Here we demonstrate a single-mode, vector vortex beam exciton-polariton laser from a monolithic vertical cavity, operating in the strong-coupling regime below band inversion density. 

Exciton-polariton lasers are a type of coherent light source with low power threshold, as they rely on the stimulated scattering and condensation of polaritons---hybrid light-matter quasi-particles in microcavities---below the band inversion density \cite{deng_exciton-polariton_2010}. 
Due to local defects and polariton nonlinearity, rich spin and phase structures can form in a polariton condensate, which are directly transferred to the emitted photons, producing a variety of vortices in the emission \cite{lagoudakis_quantized_2008, roumpos_single_2011, manni_hyperbolic_2013}. 
In particular, a polariton condensate with a spin vortex can emit vector vortex beams. 
Utilizing the intrinsic but relatively weak polarization anisotropy of the exciton-photon coupling in cavities made of distributed Bragg reflectors, lasing in different spin vortex states has been demonstrated, when the cavity was etched into coupled-circular pillars or formed with an external concave dielectric mirror \cite{sala_spin-orbit_2015, dufferwiel_spin_2015}. 
However, since the anisotropy is weak, selection of a specific vortex state required careful tuning of the optical pump \cite{sala_spin-orbit_2015} or external spectral filtering \cite{dufferwiel_spin_2015}. 

In this work, we demonstrate robust, single-mode vector vortex beam polariton lasing, in a compact monolithic vertical cavity. 
It does not require special pumping schemes, spectral filtering, or other external mechanisms to select among different spin vortex states. 
Instead, the lasing mode is determined by the structure of the microcavity. 
This is achieved by using a concentric circular dielectric sub-wavelength grating (SWG) in lieu of a flat mirror \cite{kim_monolithic_2019} to enable control of the polarization vector of the cavity mode. 
Polariton lasing in radially polarized spin vortex state is experimentally observed. 
Polarization of the vector vortex beam emission is measured and Stokes parameters are calculated to verify the radial polarization distribution. 
Polarization-resolved spatial interferometry confirms that the state is a coherent superposition of the two circular polarization components with opposite phase vortex. 
This work demonstrates a simple and scalable way to design and control the polarization properties of the polariton laser for efficient, on-chip vector vortex beam lasers.

\section{Setup and Principle of the Experiment}

\begin{figure}[htbp]
\begin{center}
\includegraphics{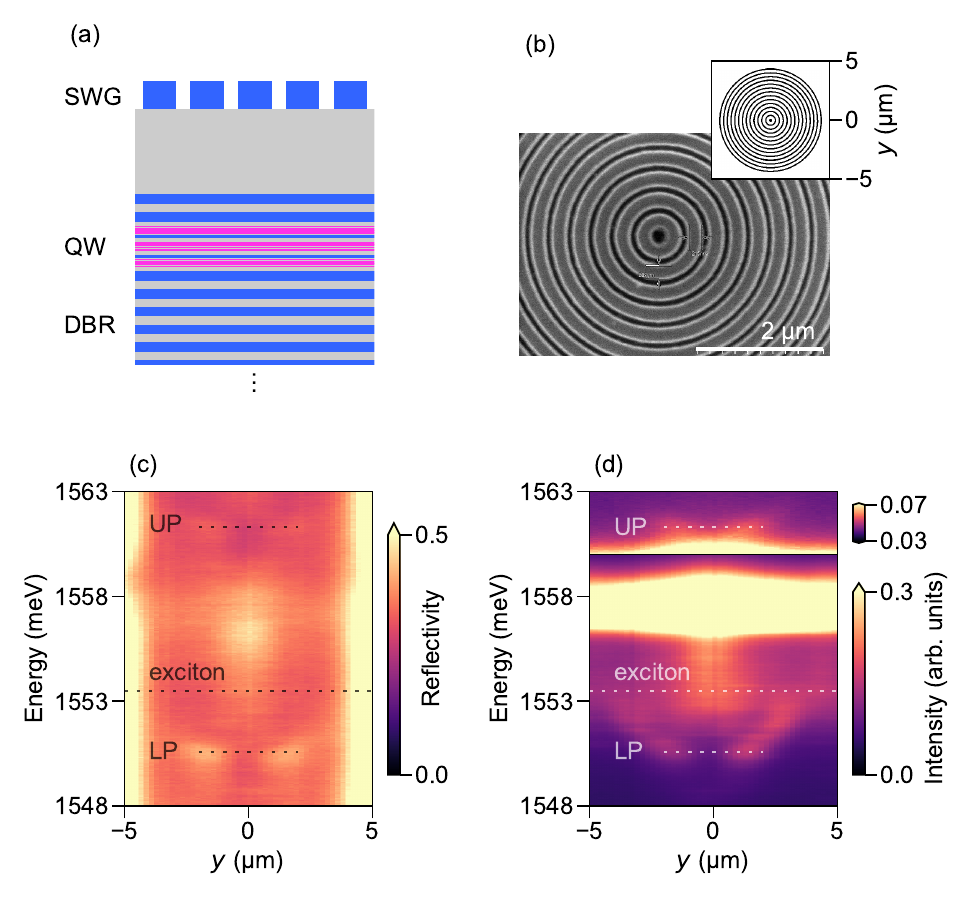}
\caption{\textbf{Polariton laser device with concentric circular grating mirror.} 
(a) Schematic of the SWG-based microcavity, showing the SWG-based top mirror, QWs in cavity and part of the bottom DBR mirror. 
(b) Top view SEM image of a fabricated device. 
Inset: design of the concentric circular pattern. 
The black regions are the etched grating grooves. 
(c) Real-space reflection spectrum across the center of the device, $y$-polarized. 
Here, the light is incident from the air side of the grating, where most energy goes into first diffraction order of transmission \cite{kim_monolithic_2019}. 
The reflectivity is much lower than 1 and the resonances appear as peaks. 
(d) Real-space PL spectrum at low excitation power with emission intensity normalized to the maximum value in this spectrum, $y$-polarized. 
In (c) and (d) the dashed lines mark the energy of the exciton resonance that strongly-couples to the cavity, and the lower polartiton (LP) and upper polariton (UP) energies about $\SI{1}{\mu m}$ away from the center of the device.}
\label{fig:device}
\end{center}
\end{figure}

A schematic of the SWG-based microcavity is shown in Fig.~\ref{fig:device}a. 
It has a $\frac{1}{2}\lambda$ $\mathrm{AlAs}$ cavity and three sets of four $\SI{12}{nm}$-wide $\mathrm{GaAs}$ quantum wells (QWs) at each of the three center antinodes of electric field. 
The bottom mirror is a regular distributed Bragg reflector (DBR) with 37-pairs of $\mathrm{AlAs}$-$\mathrm{Al_{0.15}Ga_{0.85}As}$ layers. 
The top mirror consists of a short 2.5-pair $\mathrm{AlAs}$-$\mathrm{Al_{0.15}Ga_{0.85}As}$ DBR, followed by a $\SI{600}{nm}$ $\mathrm{Al_{0.85}Ga_{0.15}As}$ spacer layer, and a $\SI{200}{nm}$ thick $\mathrm{Al_{0.15}Ga_{0.85}As}$ layer etched into an SWG. 
The short top DBR has a low reflectance. Its main function is not to be a reflector, but to embed and protect one set of the QWs. 
The spacer layer thickness is adjusted to ensure round-trip phase matching for the cavity mode. 
The SWG is the key component that provides both high reflectivity and high mode selectivity. 

It has been shown that an SWG made of straight, periodic bars can function as a polarization-selective mirror with high enough reflectance to support polariton lasing \cite{zhang_zero-dimensional_2014, kim_monolithic_2019}. 
This is because the guided wave resonances in the grating are polarized parallel or perpendicular to the grating; they can be designed to interfere at the surface of the grating to allow nearly perfect reflectance for one of the polarizations, while the orthogonally polarized mode has low reflectance \cite{lalanne_optical_2006, chang-hasnain_high-contrast_2012, magnusson_wideband_2014, gebski_monolithic_2015}. 

While the above principle is based on straight gratings, here we extend it to curved gratings to allow control of the polarization distribution of the cavity mode. 
Specifically, to create a vector vortex beam, we ``bend'' the straight gratings into concentric circular pattern (Fig.~\ref{fig:device}b) with diameters up to $\SI{8.65}{\mu m}$. 
The grating parameters are optimized to have high reflectivity only for light polarized perpendicular to the grating bars (transverse-magnetic, TM) incident from the cavity \cite{kim_monolithic_2019}, including a grating period of about $\SI{336}{nm}$ and a duty cycle, defined as the ratio of the width of the high-index material to the period, of about $0.7$.  
The orthogonal polarization (transverse-electric, TE) has low reflectivity. 
With the circular pattern, the TM mode becomes polarized radially around the center, forming the desired vector vortex mode.  
The angular dependence of polarization of a radially polarized state can be described by a superposition of two orthogonal circular polarizations, each with a phase vortex whose cores are at the same location but have opposite phase winding direction:
\begin{equation}\label{eq:radial_spin_vortex}
\psi = e^{-i\theta}\begin{bmatrix}1\\i\end{bmatrix} + e^{i\theta}\begin{bmatrix}1\\-i\end{bmatrix},
\end{equation}
where $\theta$ is the in-plane azimuthal angle and $\left[1,i\right]^T$ and $\left[1,-i\right]^T$ represent the counter-clockwise and clockwise circular polarization components respectively. 

The reflectance of this vector vortex mode remains high despite the ``bending'' of the grating. 
This is because, away from the center, where the radius of curvature is larger than the period, the SWG can be approximated as straight with high local reflectance for TM-polarized field. 
Near the center of the device, the local reflectance will be lower, but the field intensity of the vector vortex mode diminishes. 
Therefore, the overall reflectance for the vector vortex mode remains high, allowing strong-coupling with excitons and formation of vector vortex polariton modes. 

Such a cavity is well-suited for achieving single-mode vector vortex polariton lasing for its high mode selectivity. 
Although photon modes of different polarization distributions are also allowed in the cavity \cite{dufferwiel_spin_2015}, as there is a family of vector vortex beams that are paraxial solutions to the free space vector electromagnetic wave equation \cite{zhan_cylindrical_2009}, only the radially polarized modes have high enough Q-factor to form polaritons via strong-coupling. 
Other modes remain in the weak-coupling regime and have much higher dissipation rates. 
Due to the lower threshold than conventional photon lasing, polariton condensation is expected as the mechanism of laser emission in this device. 
The polarization structure of the laser emission will be the same as the polaritons---radially polarized. 

To further facilitate polariton condensation, we utilize the design flexibility of SWG to engineer a trapping potential landscape \cite{balili_role_2009}. 
The period and duty cycle of the grating can be tuned to change the reflection phase of the grating, and consequently change the resonance energy of the cavity \cite{kim_monolithic_2019}. 
Changing the duty cycle of the grating along the radial direction, from $0.7$ at the center to $0.65$ at the edge, we create a circular trapping potential of about $\SI{2}{meV}$ for the polaritons. 

\section{Result and Analysis}

We first characterize the polariton modes by reflection (Fig.~\ref{fig:device}c) and photoluminescence (PL) spectroscopies at low excitation powers (Fig.~\ref{fig:device}d). 
Lower and upper polariton branches are visible in the spectra, with energies $\SI{1550.6}{meV}$ and $\SI{1561.3}{meV}$ at about $\SI{1}{\mu m}$ away from the center of the device.  
As a result of the modulation of the grating duty cycle, the energies of the polariton branches are increasing from near the center to the boundary of the device. 
On the unetched part of the sample, we measure two exciton resonances at $\SI{1553.5}{meV}$ and $\SI{1557.2}{meV}$. 
These are due to inhomogeneity among the quantum wells \cite{kim_monolithic_2019}. 
The exciton resonance at $\SI{1553.5}{meV}$ is coupled to the cavity, forming the polariton modes. 
The remaining exciton resonance at $\SI{1557.2}{meV}$ is still visible in the spectra of the device. 
The estimated Rabi splitting, defined as the energy difference between the two polariton branches when the cavity is in resonance with the exciton, is $\SI{9.5}{meV}$ and the cavity-exciton detuning is about $\SI{4.9}{meV}$ near the center of the device. 

\begin{figure}[htbp]
\begin{center}
\includegraphics{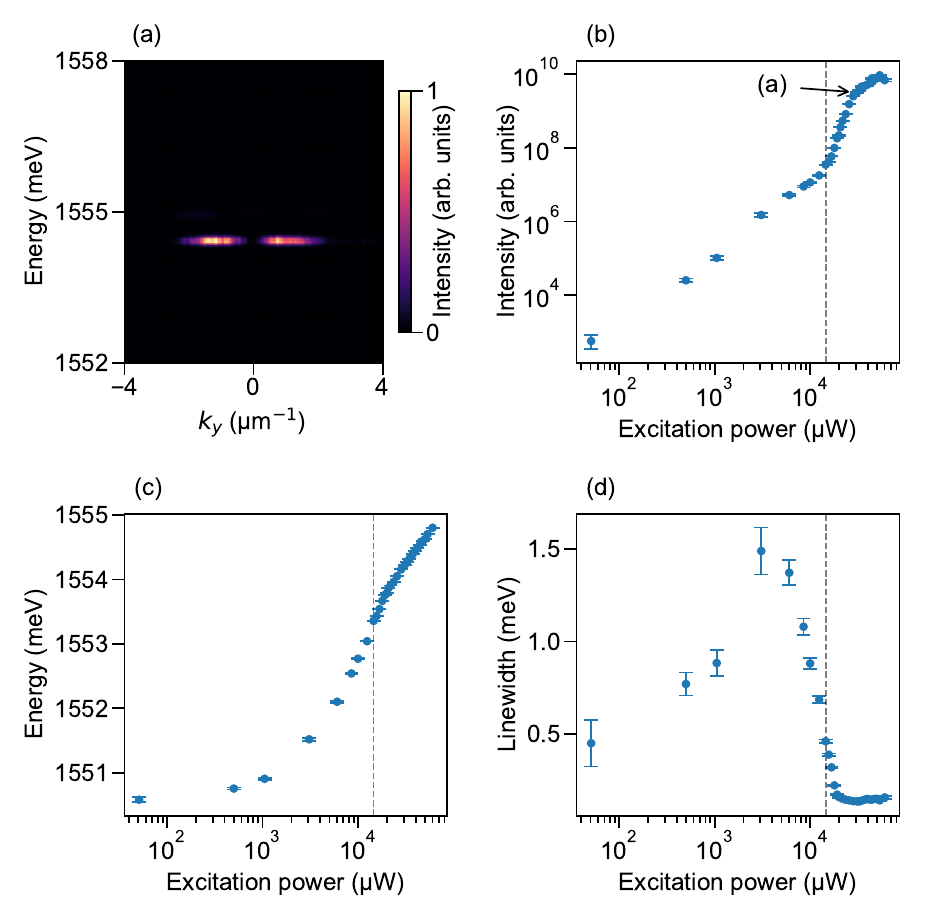}
\caption{\textbf{Ground state lasing properties.} 
(a) $k$-space emission spectrum at $\SI{30.8}{mW}$ excitation power. 
The intensity is normalized to the maximum value in this spectrum. 
(b) Emission intensity. 
(c) Energy. 
(d) Linewidth. 
In (b-d) the gray dashed line marks the lasing threshold.}
\label{fig:lasing}
\end{center}
\end{figure}

We pump the sample using a non-resonant continuous-wave laser at $\SI{784}{nm}$. 
The illuminated area is extended to $\SI{10}{\mu m}$ in diameter with an additional lens so that it covers the whole device. 
We measure the excitation power dependence of the reciprocal-space spectra of the device, with a linear polarizer (in $y$-direction) in the collection optical path. 
Since the direction of the grating bars changes within the device, emission from both the strong-coupling polarization and the weak-coupling polarization are collected. 
With increasing excitation power, the polariton ground state, which exhibits a two-lobe distribution in $k$-space, becomes the most populated (Fig.~\ref{fig:lasing}a). 
The intensity of the emission from this state is shown in Fig.~\ref{fig:lasing}b, which is obtained by first applying curve fitting to the spectra at $k_y=\pm \SI{0.9}{\mu m^{-1}}$ to separate the emission from the polariton state and the broad exciton peaks, and then calculating the area of the peak corresponding to the polariton state. 
A threshold can be observed at $\SI{14.5}{mW}$ excitation power. 
The energy of the polariton ground state increases continuously, both before and after the threshold (Fig.~\ref{fig:lasing}c), and remains well below the cavity resonance of $\SI{1558.4}{meV}$. 
The linewidth (Fig.~\ref{fig:lasing}d) is first broadened up to $\SI{1.5}{meV}$ before threshold due to increased scattering, and then quickly narrows down to below the spectral resolution of $\SI{0.2}{meV}$ of our instrument.  

To measure and characterize the polarization of the emission, the device under excitation is imaged with a linear polarizer and a quarter wave-plate. 
The intensity distribution of the vertical, horizontal, diagonal, and circular polarizations of the emission can be measured separately. 
The Stokes parameters are then calculated by
\begin{equation}
S_1 = \frac{I_H-I_V}{I_H+I_V},\quad S_2 = \frac{I_{D+}-I_{D-}}{I_{D+}+I_{D-}},\quad S_3 = \frac{I_{\sigma+}-I_{\sigma-}}{I_{\sigma+}+I_{\sigma-}},
\end{equation}
where $I_H$ and $I_V$ are the intensities of vertical and horizontal, $I_{D+}, I_{D-}$ the two diagonal, and $I_{\sigma+}, I_{\sigma-}$ the two circular polarization respectively. 
The degree of polarization (DOP) is calculated by $p = \sqrt{S_1^2+S_2^2+S_3^3}$. 

\begin{figure}[htbp]
\begin{center}
\includegraphics{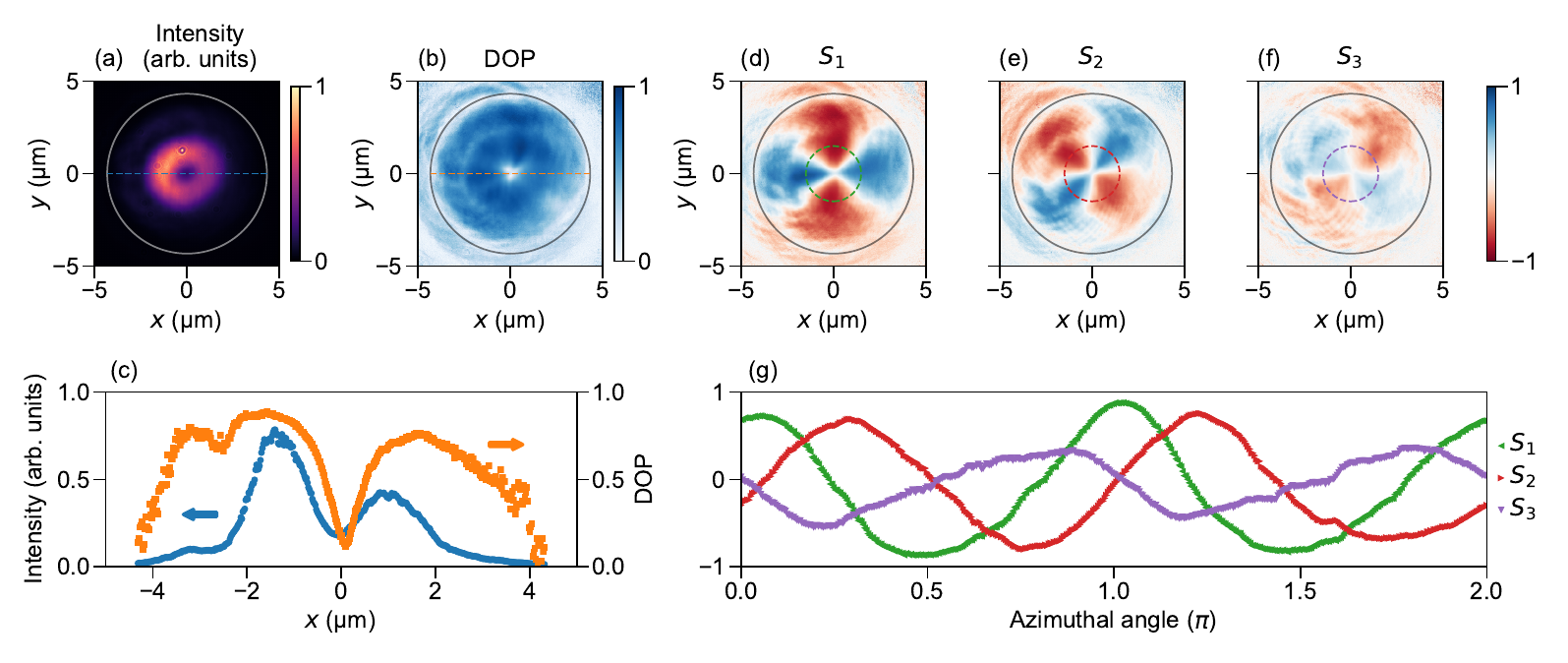}
\caption{\textbf{Polarization properties of the emission at $\SI{40}{mW}$ excitation power.} 
Spatial distribution of (a) normalized emission intensity and (b) degree of polarization. 
(c) Profiles of total intensity (blue circles) and DOP (orange squares) along $y=0$ (marked with dashed lines in (a) and (b)). 
(d-f) Stokes parameters $S_1$-$S_3$. 
(g) Azimuthal profiles of $S_1$-$S_3$ at $r=\SI{1.5}{\mu m}$ (marked with dashed circles in (d-f)). 
In (a-b) and (d-f) the solid gray circles mark the boundary of the device. 
In (a) and (c) the intensity is normalized to the maximum value in this image.}
\label{fig:polarization}
\end{center}
\end{figure}

As an example, at $\SI{40}{mW}$ excitation, which is well above threshold, the emission within the device region is highly polarized as shown in Fig.~\ref{fig:polarization}b, c. 
The peak DOP is above $0.9$, while at some spatial region where the ground state intensity is weaker, the emission from other states could lower the DOP. 
Comparing $S_1$ (Fig.~\ref{fig:polarization}d, g) and $S_2$ (Fig.~\ref{fig:polarization}e, g) with $S_3$ (Fig.~\ref{fig:polarization}f, g), it is clear that the emission is dominantly polarized perpendicular to the grating bars, or, the polarization is rotating and always in the radial direction as the azimuthal angle changes around the device. 
This confirms polariton lasing in a radially polarized spin vortex state. 

To verify that the polarization properties are from the coherent emission of one state spanning the whole device, we measure by interferometry the phase structure of the emission in different polarization basis. 
For the radially polarized spin vortex, the phase structure of any chosen polarization component should agree with Eq.~\ref{eq:radial_spin_vortex}. 
However, if the emission is from independent parts at different regions of the device, the phase structures described by Eq.~\ref{eq:radial_spin_vortex} are not expected and the interferogram may not even have visible fringe patterns over the scale of the whole device. 

The emission beam, after polarization filtering, is divided into two paths and then overlapped and made to interfere with a relative spatial displacement. 
In this arrangement, the center of the emission pattern, expected to have phase singularities, can interfere with the part of the emission that has continuous phase, giving rise to fringe patterns characteristic of these phase singularities. 
We apply off-diagonal Fourier filtering to the interferograms, isolating the first order Fourier component, which contains the information of the phase difference between the two paths \cite{manni_hyperbolic_2013}. 

\begin{figure}[htbp]
\begin{center}
\includegraphics{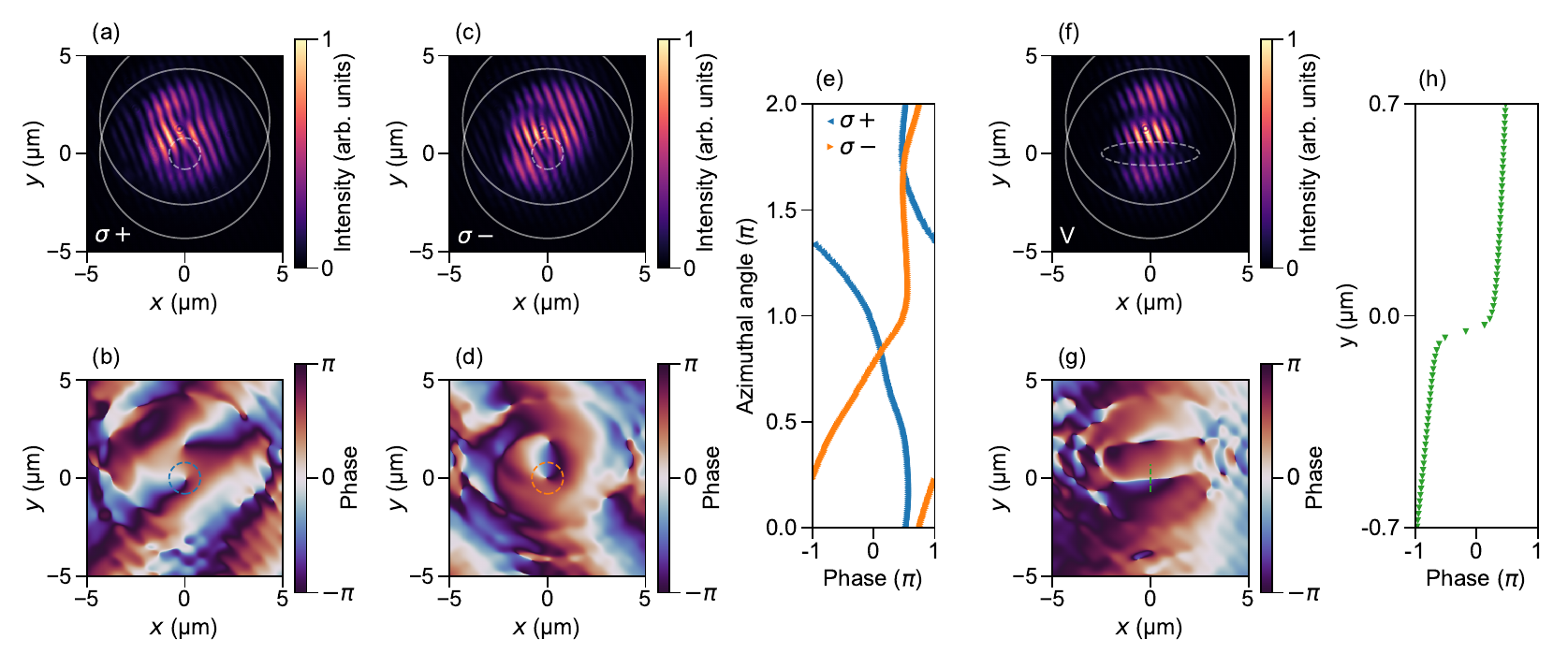}
\caption{\textbf{Polarization-resolved interferograms at $\SI{40}{mW}$ excitation power.} 
(a) Interferogram of $\sigma+$ component, with the extracted phase map in (b). 
The two solid circles mark the device boundary in the two interfering images and the dashed circle marks one of the fork patterns. 
(c, d) Similar to (a, b) but for $\sigma-$ polarization. 
(e) Azimuthal profiles of the phase at $r=\SI{0.8}{\mu m}$ (dashed circles in (b) and (d)) 
(f, g) Similar to (a, b) but for vertical linear polarization. 
The dashed ellipse in (f) marks the dislocation of fringes. 
(h) Phase profile along the dashed line in (g). 
In (a), (c), and (f) the intensity is normalized to the maximum value in the corresponding image.}
\label{fig:interferogram}
\end{center}
\end{figure}

The measured interferograms of each of the circular polarization components show two fork-like patterns (Fig.~\ref{fig:interferogram}a, c), which are signatures of the interference of a phase vortex with continuous phase \cite{lagoudakis_quantized_2008}, aligned along the direction of displacement between the two paths. 
Furthermore, for the two circular polarizations, the forks open towards opposite direction, as expected from the opposite sign of the phases vortices. 
The phase singularities and their opposite direction of phase winding is clearly seen in the extracted phase map in Fig.~\ref{fig:interferogram}b, d, e. 
For any linear polarization, the emission should have a sinusoidal azimuthal distribution, which translates to a real space distribution of two lobes of $\pi$ phase difference. 
Indeed, in the interferogram of the vertical polarization (Fig.~\ref{fig:interferogram}f), we observe two lines of dislocation of the fringes perpendicular to the direction of the displacement. 
The extracted phase map shows a sudden increase of $\pi$ (Fig.~\ref{fig:interferogram}g, h) across the line. 
The interference measurements corroborate the identification of the radially polarized spin vortex state. 

\section{Conclusion}

In summary, we have demonstrated a deterministic radially polarized vector vortex beam exciton-polariton laser in a relatively simple, monolithic structure, confirmed by polarization-resolved imaging and interferometry. 
With polarization-dependent high-reflectivity grating as one of the cavity mirrors, the spin texture of polaritons can be prescribed by the design of the grating pattern and no external control techniques are needed to selected the spin state. 
The device may allow low threshold power and is compatible with electrical pumping and scalable integration. 
Different designs of the grating may enable many other types of vector vortex beam lasing, such as an azimuthally polarized beam by utilizing a TE-polarized high-reflectivity mode of the grating \cite{zhang_zero-dimensional_2014}, or a hyperbolic or spiral vortex beam with a grating pattern different from the concentric circles.

\begin{acknowledgments}
J.H., S.K. and H.D. acknowledge financial support from the US Air Force Office of Scientific Research under grant FA9550-15-1-0240 and the US National Science Foundation (NSF) under grant DMR 1150593 and DMR 2004287. 
The Würzburg group gratefully acknowledges support by the state of Bavaria. 
\end{acknowledgments}

\end{document}